\documentclass{book}
\usepackage{Wires}
\usepackage[dvips,pdfmark]{hyperref}
\usepackage{endfloat}

\DeclareMathOperator{\argmin}{arg\,min}
\DeclareMathOperator{\E}{E}

\begin{document}
\AdvancedReview

\arttitle{Semi-supervised clustering methods}

\artid{2CAC}

\fname{Eric}
\sname{Bair}

\affil{Departments of Endodontics and Biostatistics \\
  Univ. of North Carolina at Chapel Hill \\
  Chapel Hill, NC 27599}


\begin{abstract}
Cluster analysis methods seek to partition a data set into homogeneous
subgroups. It is useful in a wide variety of applications, including
document processing and modern genetics. Conventional clustering
methods are unsupervised, meaning that there is no outcome variable
nor is anything known about the relationship between the observations
in the data set. In many situations, however, information about the
clusters is available in addition to the values of the features. For
example, the cluster labels of some observations may be known, or
certain observations may be known to belong to the same cluster. In
other cases, one may wish to identify clusters that are associated
with a particular outcome variable. This review describes several
clustering algorithms (known as ``semi-supervised clustering''
methods) that can be applied in these situations. The majority of
these methods are modifications of the popular k-means clustering
method, and several of them will be described in detail. A brief
description of some other semi-supervised clustering algorithms is
also provided.
\end{abstract}

\begin{introtext}
The objective of cluster analysis is to partition a data set into a
group of subsets (i.e. ``clusters'') such that observations within a
cluster are more similar to one another than observations in other
clusters. For a more detailed discussion, see \citet{HTF09} or
\citet{aG99}.

Traditional clustering methods are unsupervised, meaning that there is
no outcome measure and nothing is known about the relationship between
the observations in the data set. However, in many situations one may
wish to perform cluster analysis even though an outcome variable
exists or some preliminary information about the clusters is known.
For example, an e-mail classification procedure may seek to
characterize the properties of ``spam'' e-mails. Suppose a large
database of e-mails is available, a small subset of which has already
been classified as ``spam'' or ``not spam.'' One may wish to identify
clusters in this data set such that one cluster consists primarily of
``spam'' and the other cluster consists primarily of ``not spam.'' Or
in a genetic study of cancer, one may wish to identify genetic
clusters that can be used to determine the prognosis of cancer
patients. Such clusters would only be of interest if they were
associated with the outcome of interest, namely patient survival.

Clustering methods that can be applied to partially labeled data or
data with other types of outcome measures are known as semi-supervised
clustering methods (or sometimes as supervised clustering
methods). They are examples of semi-supervised learning methods, which
are methods that use both labeled and unlabeled data \citep{BM98,
tJ99, kN00, sB04b}. This review will briefly describe several
semi-supervised clustering methods that can be applied to different
types of partially labeled data sets. The review will focus primarily
on variations of k-means clustering, since most existing
semi-supervised clustering methods are modified versions of k-means
clustering. However, a brief description of some semi-supervised
hierarchical clustering methods will also be provided.
\end{introtext}

\section{Traditional (Unsupervised) Clustering Methods}
This section will briefly describe two of the most common traditional
cluster analysis methods, namely k-means clustering and hierarchical
clustering.

\subsection{K-Means Clustering}
K-means clustering is one of the most popular cluster analysis
methods. It is generally applied to data sets where all the variables
are quantitative and the distance between observations is measured
using the squared Euclidean distance, which is defined as follows:
\begin{equation} \label{E:euc_dist}
  d(x_i, x_{i'}) = \sum_{j=1}^p (x_{ij} - x_{i'j})^2
\end{equation}
Here $x_i$ and $x_{i'}$ are observations from a data set with $p$
features, and $x_{ij}$ represents the value of the $j$th feature for
observation $i$. The k-means clustering algorithm attempts to assign
each observation to a cluster to minimize the following objective
function:
\begin{equation} \label{E:km_obj_func}
  \sum_{k=1}^K \sum_{C_i=k} \sum_{C_{i'}=k} \sum_{j=1}^p
  (x_{ij}-x_{i'j})^2
\end{equation}
In the above expression, $K$ represents the number of clusters, and
$C_i$ represents the cluster to which observation $i$ is assigned,
where $1 \leq C_i \leq K$. This objective function is also known as
the ``within-cluster sum of squares'' or WCSS. Note that
(\ref{E:km_obj_func}) can be written as:
\[
\sum_{k=1}^K n_k \sum_{C_i=k} \sum_{j=1}^p (x_{ij} - \bar{x}_{kj})^2
\]
where $n_k$ is the number of observations in cluster $k$ and
$\bar{x}_{kj}$ is the mean of feature $j$ in cluster $k$.

Several k-means clustering algorithms have been proposed to minimize
(\ref{E:km_obj_func}) \citep{eF65, jM67, HW79, sL82}. However, each
algorithm uses some variation of the following strategy:

\begin{enumerate}
\item Randomly assign each observation to an initial cluster.
\item For each feature $j$ and cluster $k$, calculate $\bar{x}_{kj}$,
  the mean of feature $j$ in cluster $k$. \label{en:calc_mean}
\item Assign each observation $i$ to a new cluster $C_i$ as follows:
  \[
  C_i = \argmin_k \sum_{j=1}^p (x_{ij}- \bar{x}_{kj})^2
  \]
  \label{en:update_clust}
\item Repeat steps \ref{en:calc_mean} and \ref{en:update_clust} until
  the algorithm converges.
\end{enumerate}

The above algorithm is guaranteed to converge, but it may converge to
a local minimum. Hence, it is advisable to repeat the algorithm
multiple times with different initial clusters and choose the set of
clusters that produces the minimum WCSS. For a more detailed
discussion of k-means clustering and several variations of the k-means
algorithm see \citet{HTF09}.

The k-means clustering algorithm requires one to choose the number of
clusters $K$. Several methods have been proposed for choosing $K$. One
common method is the ``gap statistic'' proposed by \citet{TWH01}. Let
$W_k$ be the WCSS (\ref{E:km_obj_func}) when $K=k$. It is simple to
verify that $W_k$ will always decrease as $k$ increases, so one cannot
simply choose the value of $K$ that minimizes $W_K$. The motivation
for the gap statistic is the following: Let $K^*$ denote the true
value of $K$. If $k<K^*$, then at least one cluster produced by the
k-means is actually two separate clusters, and so $W_{k+1}$ should be
significantly smaller than $W_k$. On the other hand, if $k>K^*$, then
at least two clusters produced by the k-means algorithm are actually a
single cluster, so $W_{k-1}$ should be only slightly larger than
$W_k$. Thus, the gap statistic seeks to identify the smallest $K$ such
that $W_k$ does not decrease significantly for $k>K$.

Formally, the gap statistic is defined to be
\[
  G_k = \E \left[\log(W_k)\right] - \log(W_k)
\]
The expected value $\E \left[\log(W_k)\right]$ is calculated under a
suitable reference distribution. One common choice of a reference
distribution is a multivariate uniform distribution with the same
range as the data set of interest. In this case, this expected value
may be estimated by sampling from this (uniform) reference
distribution. \citet{TWH01} estimate the number of clusters $K$ as
follows:
\[
  \hat{K} = \argmin_K \left\{K|G_K \geq G_{K+1} - s_{K+1} \right\}
\]
where $s_k$ is the estimated standard deviation of $\E
\left[\log(W_k)\right]$. The idea is that when $k \geq K^*$ then
$G_{k+1} \approx G_k$, so one may estimate $K^*$ by choosing the
minimum $k$ such that $G_{k+1} \approx G_k$.

A number of other methods have been proposed for choosing the number
of clusters $K$ \citep{MC85, SJ03, TW05}. See the aforementioned
references for details of these methods.

\subsection{Hierarchical Clustering}
K-means clustering is an example of what are known as partitional
clustering methods, which partition a data set into a fixed number of
disjoint subgroups. In contrast, hierarchical clustering groups data
points into a series of clusters in a tree-like structure. At each
level of the tree, clusters are formed by merging clusters at the next
lower level of the tree. Thus, each data point forms a singleton
cluster at the bottom level of the tree, and the top level of the tree
consists of a single cluster containing all of the data points.

There are a wide variety of different methods for hierarchical
clustering. This review will briefly describe a few of the most common
hierarchical clustering methods, although many other hierarchical
clustering methods have been proposed. See \citet{HTF09} for more
information (including descriptions of several other hierarchical
clustering methods).

\begin{figure}
  \centerline{\includegraphics[width=0.5\textwidth]{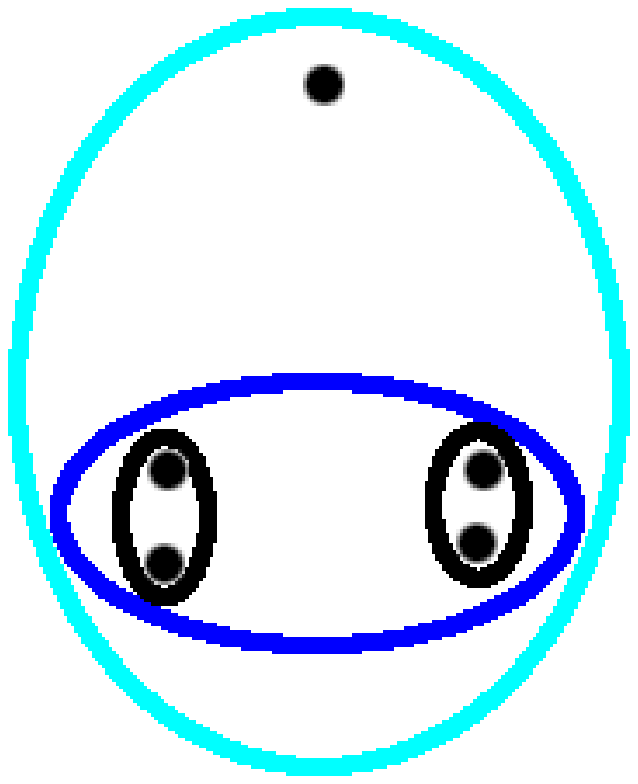}}
  \caption[Illustration of hierarchical clustering]{This figure
    illustrates how hierarchical clustering would partition a simple
    data set. In the first two steps, the two pairs of adjacent points
    would each be combined into a single cluster. In the third step,
    these two clusters would be combined into a larger cluster. In the
    final step, the remaining point would be combined to this
    cluster. All the data points are now combined into a single
    cluster, so the algorithm terminates.}
  \label{F:hier_clust}
\end{figure}

One of the most common hierarchical clustering methods is
agglomerative hierarchical clustering. Agglomerative hierarchical
clustering methods start with the set of individual data points and
merge the two ``most similar'' points into a cluster. At each step of
the procedure, the two ``most similar'' clusters (which may be
individual data points) are merged until all of the data points have
been merged into a single cluster. See Figure \ref{F:hier_clust} for
an illustration of agglomerative hierarchical clustering.

In order to apply the hierarchical clustering algorithm described
above, one must define how the pair of ``most similar'' clusters is
chosen. Note that for hierarchical clustering it is not sufficient to
define a dissimilarity (or distance) measure between pairs of points;
one must also define a dissimilarity measure between pairs of
clusters. Many different dissimilarity measures have been proposed for
hierarchical clustering, but the most commonly used methods start by
defining a dissimilarity measure between pairs of points. The
Euclidean distance defined in (\ref{E:euc_dist}) is a common choice,
but other dissimilarity measures are possible. For example, when
clustering DNA microarray data, is it common to define the
dissimilarity measure between two points to be $1-\rho$, where $\rho$
is the Pearson correlation coefficient between the two
points. \citep{mE98}

Once a dissimilarity measure between two points has been defined,
there are several ways to define distances between clusters. Two
common dissimilarity measures are known as ``single linkage'' and
``complete linkage.'' Let $C_1$ and $C_2$ denote the indices of the
elements in two clusters. In other words, $i \in C_1$ if and only if
data point $x_i$ is contained in the first cluster. Also, let $d(x_i,
x_{i'})$ be the dissimilarity between data points $x_i$ and
$x_{i'}$. Then the single linkage dissimilarity between the two
clusters is defined
to be
\[
  d(C_1, C_2) = \min_{i \in C_1,\, i' \in C_2} d(x_i, x_{i'})
\]
and the complete linkage dissimilarity is defined to be
\[
  d(C_1, C_2) = \max_{i \in C_1,\, i' \in C_2} d(x_i, x_{i'})
\]
Other dissimilarity measures between clusters can also be used. For
example, one could define the dissimilarity between two clusters to be
the average dissimilarity between the elements of the two clusters:
\[
  d(C_1, C_2) = \frac{1}{n_1 n_2} \sum_{i \in C_1} \sum_{i' \in C_2}
  d(x_i, x_{i'})
\]
where $n_1$ and $n_2$ are the number of data points in clusters 1 and
2, respectively. Each such dissimilarity measure between clusters has
certain advantages and disadvantages. See \citet{HTF09} for details.

\begin{figure}
  \centerline{\includegraphics[width=\textwidth]{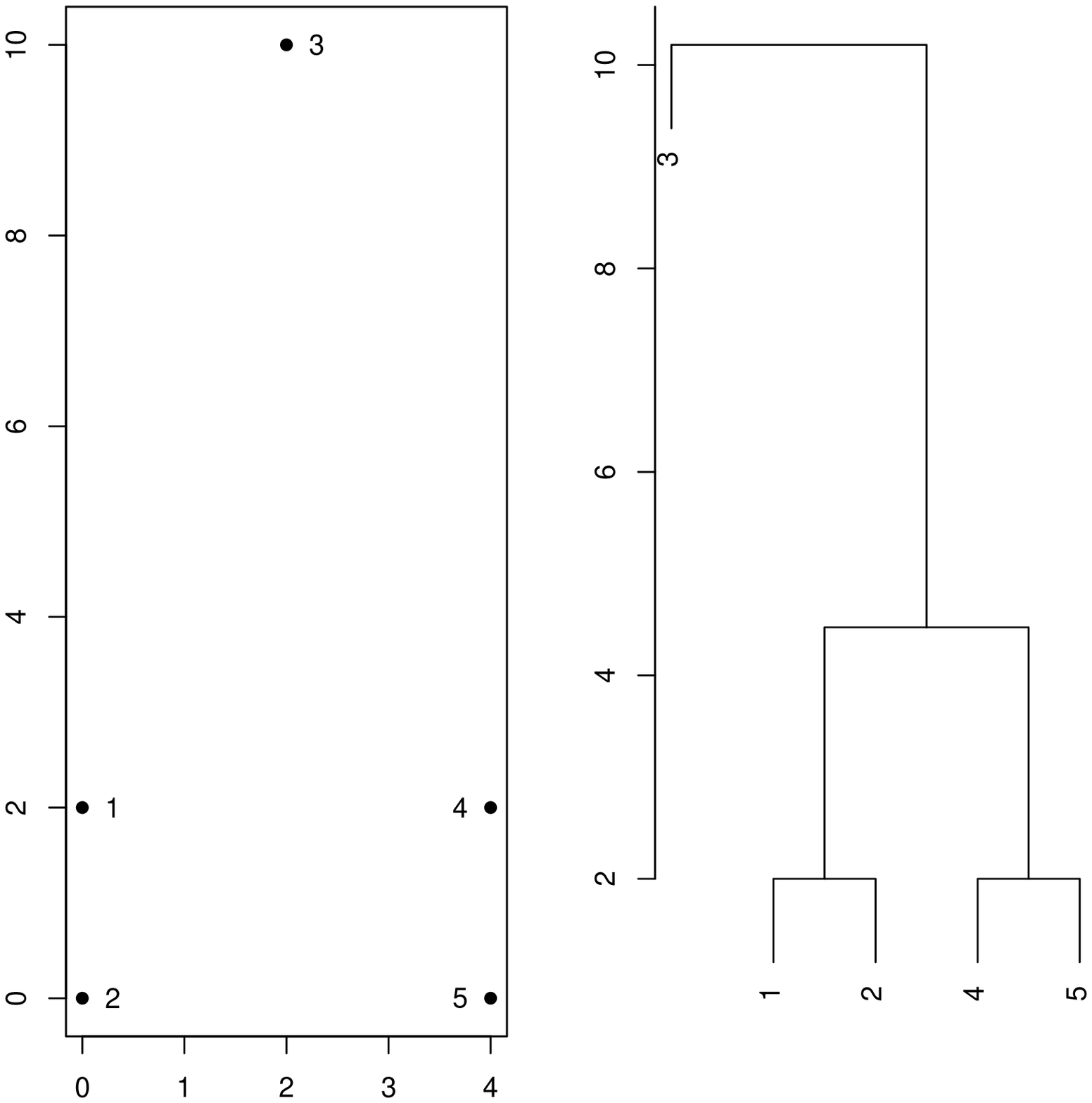}}
  \caption[Example of a cluster dendogram]{Hierarchical clustering was
    applied to the five data points plotted in the left panel. The
    resulting dendogram is shown on the right panel. Note that point 3
    is much more distant from (and hence dissimilar to) the remaining
    four points. Thus, the height of the node where point 3 is merged
    to the remaining points is higher than the height of the other
    nodes in the graph.}
  \label{F:dendo_fig}
\end{figure}

As noted earlier, the results of hierarchical clustering may
represented as a binary tree. Each node of the tree represents a
cluster. (In particular, the root node is the top-most cluster which
contains all of the data points, and each terminal node corresponds to
a singleton ``cluster'' consisting of a single data point.) This tree
structure can be represented in a graphical form known as a
dendogram. It is customary to plot the dendogram such that the height
of each node in the tree corresponds to the dissimilarity between the
two clusters that were merged to form the cluster. See
\ref{F:dendo_fig} for an example of a dendogram of a simple data set.

\section{Semi-Supervised Clustering Methods}
We will now briefly outline several semi-supervised clustering
methods. These methods will be organized according to the nature of
the known outcome data. First, we will consider the simplest case,
namely the case where the data is partially labeled. In other words,
the cluster assignments are known for some subset of the
observations. We will then consider the case where some sort of
relationship between the features is known, and finally the case where
one seeks to identify clusters associated with a particular outcome
variable.

\subsection{Partially Labeled Data}
In some situations, the cluster assignments may be known for some
subset of the data. The objective is to classify the unlabeled
observations in the data to the appropriate clusters using the known
cluster assignments for this subset of the data.

In a certain sense, this problem is equivalent to a supervised
classification problem, where the objective is to develop a model to
assign observations in a data set to one of a finite set of classes
based on a training set where the true class labels are known.
However, traditional supervised classification methods may be
inefficient when only a small subset of the data is labeled. For
example, if one wishes to classify web pages into a discrete number of
groups, one can easily collect millions of unlabeled observations, but
classifying any given observation requires human intervention (and
hence is likely to be slow). Similarly, if one wishes to develop a
method to classify e-mails as ``spam'' or ``not spam,'' then one can
easily collect numerous unlabeled observations, but the proportion of
labeled observations will be much smaller. For these types of
problems, conventional supervised classification methods may be
inefficient since they typically do not use unlabeled data to build
the classification algorithm. Thus, the vast majority of the available
data will not be used. In these situations, one can often build more
accurate classification rules by combining both labeled and unlabeled
data. See \citet{BM98} or \citet{tJ99} for a more detailed discussion
and examples.

\citet{sB02} developed a generalization of k-means clustering (which
they called ``constrained k-means'') for the situation where class
labels are known for a subset of the observations. Once again, we let
$x_i$ and $x_{i'}$ be observations from a data set with $p$ features,
and $x_{ij}$ represents the value of the $j$th feature for observation
$i$. Suppose further that there exists subsets $S_1, S_2, \ldots, S_K$
of the $x_i$'s such that $x_i \in S_k$ implies that observation $i$ is
known to belong to cluster $k$. (Here $K$ denotes the number of
clusters, which is also assumed to be known in this case.) Let $|S_k|$
denote the number of $x_i$'s in $S_k$. Also let $S=\cup_{k=1}^K
S_k$. The algorithm proceeds as follows:

\begin{enumerate}
\item For each feature $j$ and cluster $k$, calculate the initial
  cluster means as follows:
  \[
  \bar{x}_{kj} = \frac{1}{|S_k|} \sum_{x_i \in S_k} x_{ij}
  \]
\item Assign each observation $i$ to a new cluster $C_i$. If $x_i \in
  S$, then let $C_i=S_k$, where $x_i \in S_k$. Otherwise let
  \begin{equation} \label{E:ckm_assign_clust}
    C_i = \argmin_k \sum_{j=1}^p (x_{ij}- \bar{x}_{kj})^2
  \end{equation}
  \label{en:ckm_update_clust}
\item For each feature $j$ and cluster $k$, calculate $\bar{x}_{kj}$,
  the mean of feature $j$ in cluster $k$. \label{en:ckm_calc_mean}
\item Repeat steps \ref{en:ckm_update_clust} and
  \ref{en:ckm_calc_mean} until the algorithm converges.
\end{enumerate}

Note that this procedure is identical to the conventional k-means
procedure with the exception of the initial cluster assignments (which
are generally arbitrary anyway) and step \ref{en:ckm_update_clust}. In
step \ref{en:ckm_update_clust}, labeled observations are always
assigned to their known cluster even if they are closer to the mean of
another cluster.

The constrained k-means clustering algorithm described above assumes
that none of the labeled observations are misclassified. Using the
constrained k-means clustering procedure, if a labeled observation is
misclassified, this misclassification can never be corrected, since
this observation will be assigned to the same cluster in step
\ref{en:ckm_update_clust} in every iteration of the algorithm. Thus,
\citet{sB02} recommend an alternative algorithm (which they call
``seeded k-means clustering'') that is identical to constrained
k-means clustering with the exception of step
\ref{en:ckm_update_clust}. The seeded k-means clustering algorithm
always assigns observations to the nearest cluster using
(\ref{E:ckm_assign_clust}) even if the observation is labeled. Thus,
if an observation is initially mislabeled, then the mislabeled
observation may be corrected if it is closer to the cluster center of
a different cluster.

Observe that seeded k-means clustering is identical to conventional
k-means clustering with the exception of the first step in the
procedure. Thus, seeded k-means clustering is simply conventional
k-means clustering that uses the labeled data to help choose the
initial cluster centers. A similar approach is used in the supervised
sparse clustering method of \citet{GB13}, which is described below.

Methods for clustering partially labeled data can be useful when
analyzing DNA microarray data. In a typical microarray experiment, one
measures the gene expression levels of $p$ genes for each of $n$
samples, where normally $p \gg n$. One may wish to identify clusters
of genes with similar expression levels across samples, since the
genes in each such cluster may belong to the same biological
pathway. If certain genes are known to belong to certain pathways
prior to performing the experiment, then the cluster labels for these
genes are known. In this situation, one seeks to cluster the remaining
genes using the information from the labeled genes. Several clustering
methods have been developed for the specific problem of analyzing
partially labeled microarray data \citep{mB00, aM02, jC04, QX04, zF06,
HP06, BW07, pC08, TBK09}. These methods are specifically designed for
microarray data and will not be described in this review; see the
references for details.

\subsection{Known Constraints on the Observations}
We now consider clustering when more complex relationships among the
observations are known. In particular, we will consider two types of
possible constraints among observations: ``Must-link constraints''
require that two observations must be placed in the same cluster, and
``cannot-link constraints'' require that two observations must not be
placed in the same cluster. One possible application is when repeated
measurements are collected on some subset of the experimental
units. In such a situation, one may want to assign all of the repeated
measurements of the same experimental unit to the same cluster.

Note that this is a generalization of the problem considered in the
previous section, where the cluster assignments are known for a subset
of the features. In that situation, for each feature $j$ that is known
to belong to cluster $k$, one may impose a must-link constraint
between feature $j$ and all other features known to belong to cluster
$k$ and a cannot-link constraint between feature $j$ and features
known not to belong to cluster $k$.

Numerous methods have been proposed for solving the problem of
constrained clustering. This review will briefly describe a few of the
most commonly used methods, and references for numerous other methods
are listed below. Also see \citet{BDW09} for a more detailed
description of various algorithms for constrained clustering.

\citet{kW01} proposed the following algorithm (with they called
``COP-KMEANS'') for solving clustering problems given this type of
constraint:

\begin{enumerate}
\item Randomly assign each observation to an initial cluster.
\item For each feature $j$ and cluster $k$, calculate $\bar{x}_{kj}$,
  the mean of feature $j$ in cluster $k$. \label{en:cop_calc_mean}
\item Assign each observation $i$ to a new cluster $C_i$ as follows:
  \[
  C_i = \argmin_{k \in D_{ik}} \sum_{j=1}^p (x_{ij}- \bar{x}_{kj})^2
  \]
  where
  \[
  D_{ik} = \{k: \text{no constraints are violated when observation $i$
    is assigned to cluster $k$}\}
  \] \label{en:cop_update_clust}
\item Repeat steps \ref{en:calc_mean} and \ref{en:update_clust} until
  the algorithm converges. The algorithm fails if $D_{ik} = \emptyset$
  for any $i$ at any step of the procedure.
\end{enumerate}
Note that COP-KMEANS is identical to conventional k-means clustering
with the exception of step \ref{en:cop_update_clust}. COP-KMEANS
assigns each observation to the nearest cluster such that no
constraints are violated (whereas conventional k-means clustering
assigns each observation to the nearest cluster without considering
the constraints).

One potential drawback of the COP-KMEANS algorithm is the fact that it
requires that no constraints are violated. In some situations, one may
wish to allow for the possibility that some constraints may be
violated if there is a strong evidence that a particular constraint is
incorrect. Thus, \citet{sB04} proposed a method (which they call
``PCKmeans'') that solves the problem of identifying clusters given a
set of must-link and cannot-link constraints on the observations that
allows some constraints to be violated. PCKmeans seeks to minimize a
modified version of the objective function (\ref{E:km_obj_func}) that
is defined as follows: Let observations $(x_i, x_{i'}) \in
\mathcal{M}$ if there is a must-link constraint between observations
$i$ and $i'$, and let $(x_i, x_{i'}) \in \mathcal{C}$ if there is a
cannot-link constraint between observations $i$ and $i'$. Then
PCKmeans minimizes the following objective function:
\begin{equation} \label{E:pckm_obj_func}
  \sum_{k=1}^K \sum_{C_i=k} \sum_{C_{i'}=k} \sum_{j=1}^p
  (x_{ij}-x_{i'j})^2 + \sum_{(x_i, x_{i'}) \in \mathcal{M}} l_{i,i'}
  I(C_i \neq C_{i'}) + \sum_{(x_i, x_{i'}) \in \mathcal{C}} l^*_{i,i'}
  I(C_i=C_{i'})
\end{equation}
Here $l_{i,i'}$ is a user-defined penalty for violating a must-link
constraint between observations $i$ and $i'$ and $l^*_{i,i'}$ is the
penalty for violating a cannot-link constraint between $i$ and
$i'$. See \citet{sB04} for details of the PCKmeans algorithm for
minimizing (\ref{E:pckm_obj_func}).

The methods described above modify an existing clustering method
(namely k-means clustering) such that the constraints are
satisfied. Thus, such methods are sometimes referred to as
``constraint-based methods'' in the literature \citep{sB04b,
BBM04}. In contrast, ``distance-based methods'' (or ``metric-based
methods'') use an existing clustering method but modify the metric
used to measure the ``distance'' between a pair of observations such
that the constraints are satisfied. For example, rather than using the
simple Euclidean distance (\ref{E:euc_dist}), one may use an
alternative distance metric such that two observations with a
``must-link constraint'' will necessarily have a lower distance
between them \citep{KKM02, BM03, eX03, aB03, KKM03, CY04, tL05, HK06,
LDJ07, XNZ08, WLZ08, CCM09, xY10}. Moreover, other constraint-based
methods have been proposed \citep{DR05, DR05b, LTJ05, LL05, wT07}, and
still other methods combine both of these approaches into a single
model \citep{sB04b, BBM04}. Other forms of constrained clustering are
also possible, such as clustering on graph data \citep{bK09, YO10}.
These methods will not be described further in this review; see the
original references for details.

Thus far we have also assumed that the constraints on the observations
were specified when the data was collected. In some situations, the
data analyst may have the opportunity to select some subset of the
observations and impose constraints on this subset. For example,
suppose the objective is to cluster a large set of text documents
based on the frequency of selected words that appear in the
documents. One may manually examine any given pair of documents to
determine if they should be classified to the same cluster (and hence
imposing either a must-link constraint or a cannot-link
constraint). Suppose a researcher looked up the titles of three
documents and determined that two of the documents were romance novels
for teenagers and the third document was an article from a medical
journal. In this case, the researcher would impose a must-link
constraint between the two novels and a cannot-link constraint between
each novel and the journal article. However, there is a cost
associated with making such a determination, so typically one may only
analyze a small subset of the observations. In such a situation, it is
advantageous to choose this subset to maximize the information about
the clusters.

\citet{sB04} describe a variant of PCKmeans (called ``active
PCKmeans'') that chooses a subset of the observations on which to
impose constraints such that the accuracy of the clustering algorithm
is maximized. They show that this method outperforms the generic
PCKmeans algorithm for this type of problem. For other methods for
constraint selection in this situation, see \citet{GC07} or
\citet{pM08}.

\subsection{Semi-Supervised Hierarchical Clustering}
The majority of existing semi-supervised clustering methods are based
on k-means clustering or other forms of partitional clustering.
Comparatively few semi-supervised hierarchical clustering methods have
been proposed \citep{ZL11}. This is partly due to the fact that the
problem must be formulated differently for hierarchical clustering. As
noted earlier, most semi-supervised partitional clustering methods
utilize either partially labeled data or known constraints
(e.g. ``must-link'' or ``cannot-link'' constraints) on the
observations. It is more difficult to define such constraints for
hierarchical clustering, since hierarchical clustering links all
observations in a data set at some level of the clustering
hierarchy. Thus, a ``must-link'' constraint will always be satisfied
at some level of the hierarchy and likewise a ``cannot-link''
constraint will always be violated.

Hence, semi-supervised hierarchical clustering methods have considered
different types of constraints. For example, \citet{MT10} require
observations linked by a ``must-link'' constraint to be clustered
together at the lowest possible level of the hierarchy. They further
require that observations separated by a ``cannot-link'' constraint
must not be part of the same clustering hierarchy. Thus, rather than
identifying a single clustering hierarchy, the method of \citet{MT10}
returns several clustering hierarchies. A separate hierarchy is
produced for each observation that is part of a ``cannot-link''
constraint. Several related methods have been proposed to perform
hierarchical clustering subject to such constraints \citep{DR05b,
DR09, MT10, MT11}.

Other types of constraints have been proposed for semi-supervised
hierarchical clustering. \citet{BN06} describe a method for performing
hierarchical clustering given a set of ``must-link before''
constraints, where certain a certain set of observations must be
clustered together before they are clustered with other data
points. \cite{ZL11} develop an alternative method for hierarchical
clustering with this type of constraint. \citet{ZQ10} consider
hierarchical clustering with ``ordering constraints,'' wherein
observations must be combined in a certain order. In other words,
given an ordering constraint of $(x_3, x_1, x_4, x_2)$, observations
$x_1$ and $x_3$ must be clustered together before they can be combined
into a cluster containing $x_4$, and observations $x_1$, $x_3$, and
$x_4$ must be clustered together before they can be combined into a
cluster containing $x_2$. \citet{YYS10} define ``clusterwise tolerance
based pairwise constraints'' which define ``must-link'' and
``cannot-link'' constraints between pairs of clusters based on a
weighted count of the number of such constraints that exist between
observations in the clusters. They developed algorithms for
implementing several variants of hierarchical clustering subject to
this type of constraint \citep{YYS10, YYS11, YYS12}.

Most of these methods for semi-supervised hierarchical clustering are
very new and little research has been performed on the advantages and
disadvantages of the various methods. The development of methods for
semi-supervised hierarchical clustering remains an active research
area.

\subsection{Clusters Associated with an Outcome Variable}
\begin{figure}
  \centerline{\includegraphics[width=\textwidth]{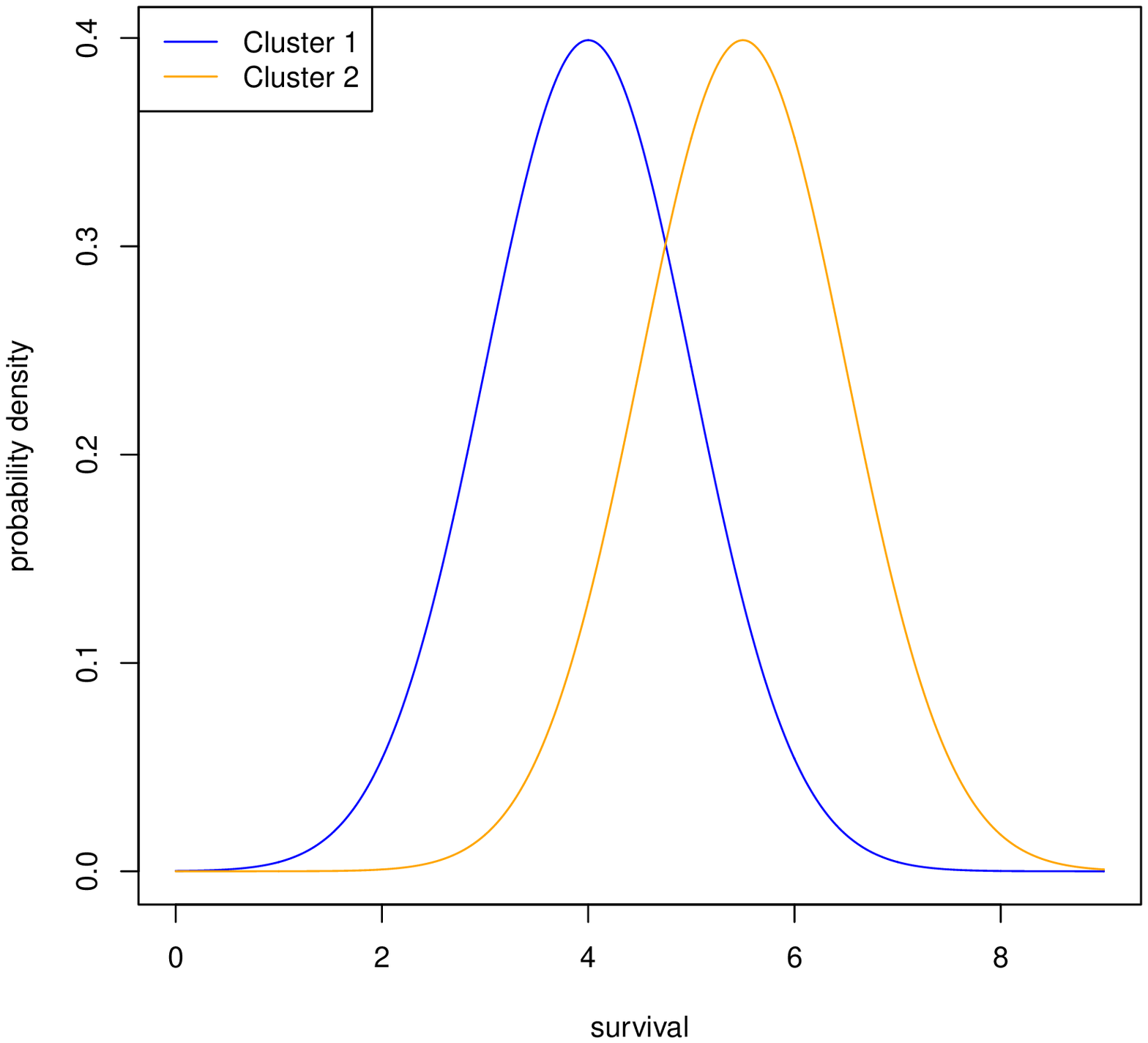}}
  \caption[An outcome variable that is a noisy surrogate for
  clusters]{This figure shows an example of a situation where an
    (observed) outcome variable (namely survival) is a ``noisy
    surrogate'' for two unobserved clusters. Suppose there are two
    subtypes of cancer, and patients with the first subtype (cluster)
    tend to have lower survival than patients with the second
    subtype. However, there is considerable overlap in the
    distribution of the survival times, so while a patient with a 
    low survival time is more likely to be in cluster 1, it is not
    possible to assign each patient to cluster based only on their
    survival time.}
  \label{F:noisy_surrogate}
\end{figure}

In other situations, one may wish to identify clusters that are
associated with a given outcome variable. Typically the outcome
variable is a ``noisy surrogate'' \citep{BT04} for the (unobserved)
clusters of interest. For example, in genetic studies of cancer, there
may exist subtypes of cancer with different genetic characteristics.
Some subtypes may be more likely to metastasize, resulting in a poorer
prognosis for patients with these subtypes. In this case these genetic
subtypes are unobserved, but the survival times of the patients in the
study may be available. A patient who has a ``high-risk'' subtype is
more likely to have a low survival time than a patient who has a
``low-risk'' subtype, but there is considerable variation within
subtypes. It is possible to observe a patient with a ``low-risk''
subtype and a low survival time (and vice versa). See Figure
\ref{F:noisy_surrogate} for an illustration of such a scenario. In this
example, patients in cluster 2 have a higher mean survival time than
patients in cluster 1, but there is significant overlap in the two
groups, so it is not possible to identify the clusters using only the
survival times.

\begin{figure}
  \centerline{\includegraphics[width=.9\textwidth]{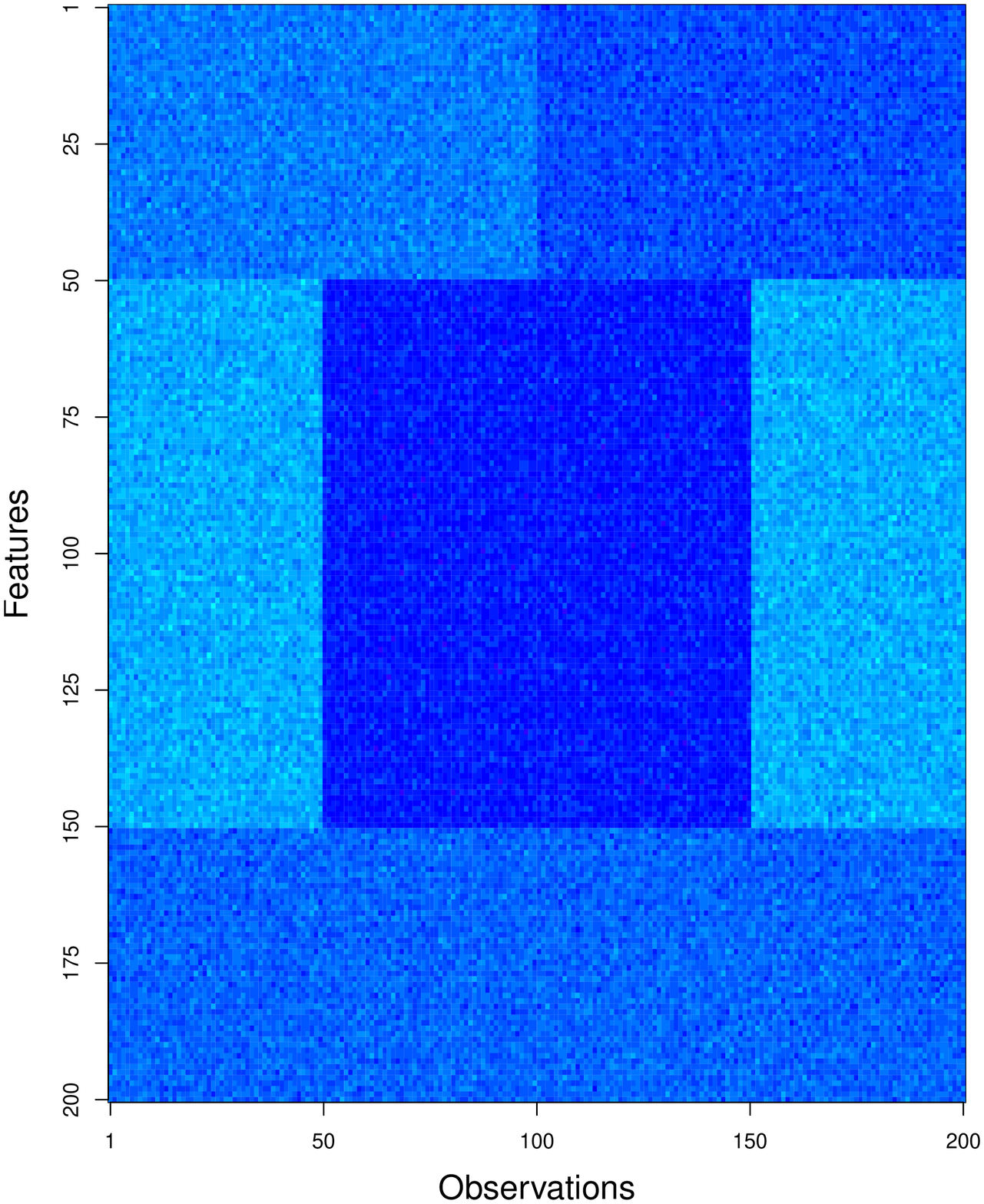}}
  \caption[Clusters that are not associated with the outcome of
  interest]{This figure shows an example of a data set where two
    different sets of clusters exist and only one cluster is
    associated with the outcome of interest. In the above figure,
    darker shades of blue correspond to higher values of the features
    and lighter shades of blue correspond to lower values. Suppose
    that observations 1-100 have a disease of interest and
    observations 101-200 are controls. In this case we would be
    interested in identifying the clusters formed by features
    1-50. However, conventional clustering algorithms will identify
    the clusters formed by features 50-150, since the distance between
    the centers of these two clusters is greater than the distance
    between the centers of the clusters formed by features 1-50.}
  \label{F:preweight}
\end{figure}

Since conventional clustering methods do not use the values of an
outcome variable, they may fail to identify clusters associated with
the outcome and instead identify clusters unrelated to the
outcome. Figure \ref{F:preweight} shows an example of a situation
where a specialized clustering method is needed to identify clusters
associated with an outcome variable of interest. In this situation,
features 1-50 form clusters that are associated with the outcome
variable and features 51-150 form clusters that are unrelated to the
outcome variable. Conventional clustering methods will nevertheless
identify the clusters defined by features 51-150, since the distance
between the centers of these clusters is greater than the distance
between the centers of the clusters defined by features 1-50. Thus,
special methods are needed to identify the clusters of interest
(i.e. the clusters defined by features 1-50) in this scenario.

Despite the importance of this problem, relatively few methods have
been proposed for identifying clusters associated with an outcome
variable. Methods exist for identifying secondary clusters for data
sets similar to the data shown in Figure \ref{F:preweight} (see for
example \citet{NT08}), but these methods also do not use information
from the outcome variable to identify the secondary clusters. One of
the earliest methods for identifying clusters associated with an
outcome variable is the ``supervised clustering'' method of
\citet{BT04}, which proceeds as follows:

\begin{enumerate}
\item For each feature in the data set, calculate a test statistic
  $T_j$ for testing the null hypothesis of no association between the
  $j$th feature and the outcome variable. If the outcome variable is
  binary (i.e. case versus control), $T_j$ may be a t-statistic. If
  the outcome variable is continuous, $T_j$ may be a t-statistic for
  testing the null hypothesis that the regression coefficient for
  predicting the outcome based on feature $j$ is equal to 0. If the
  outcome variable is a right-censored survival time, $T_j$ may be the
  corresponding test statistic from a Cox proportional hazards model.
\item Choose a threshold $M$, and apply k-means clustering to the
  features for which $|T_j|>M$. Features with $|T_j| \leq M$ are
  discarded and do not affect the cluster assignments.
\end{enumerate}

Although this approach is relatively simple, \citet{BT04} show that
this method can identify biologically relevant clusters in several
data sets. In particular, \citet{lB04} used this method to identify
subtypes of acute myeloid leukemia that were associated with patient
survival. An advantage of this method is the fact that it performs
well even when the data is high-dimensional. Since clustering is
performed using only a subset of the features, a high-dimensional data
set can be effectively reduced to a data set with fewer features.

This supervised clustering procedure requires the choice of a tuning
parameter $M$, which may be chosen using cross-validation. Also, while
the method proposed by \citet{BT04} applies k-means clustering to the
subset of the features that are most strongly associated with the
outcome variable, one could use the same strategy of selecting the
features that are most strongly associated with the outcome and then
apply hierarchical clustering or an alternative clustering
method. Indeed, \citet{dK10} propose a method called ``semi-supervised
recursively partitioned mixture models (RPMM)'' that uses this
strategy. Semi-supervised RPMM first selects a set of features that
are most strongly associated with the outcome variable and then
applies the RPMM method of \citet{eH08} to this subset of the
features. One possible advantage of RPMM over k-means clustering is
that RPMM does not require one to choose the number of clusters
$K$. \citet{dK10} provide several examples where semi-supervised RPMM
produces more accurate results than the supervised clustering method
of \citet{BT04}. However, in other situations semi-supervised RPMM can
fail to detect clusters even when such clusters exist; see
\citet{GB13} for examples.

One possible drawback to methods such as supervised clustering and
semi-supervised RPMM is the fact that any feature that is discarded
after the initial screening step is permanently excluded from the
analysis. This is problematic if one wishes to identify the features
that differ across clusters, since it is possible for features that
differ across clusters to be only weakly associated with the outcome
variable, particularly if the association between the clusters and the
outcome variable is weak. Indeed, if the association between the
clusters and the outcome variable is very weak, supervised clustering
and semi-supervised RPMM can fail to identify the correct clusters.

\begin{figure}
  \centerline{$\begin{array}{cc}
      \includegraphics[width=.5\textwidth]{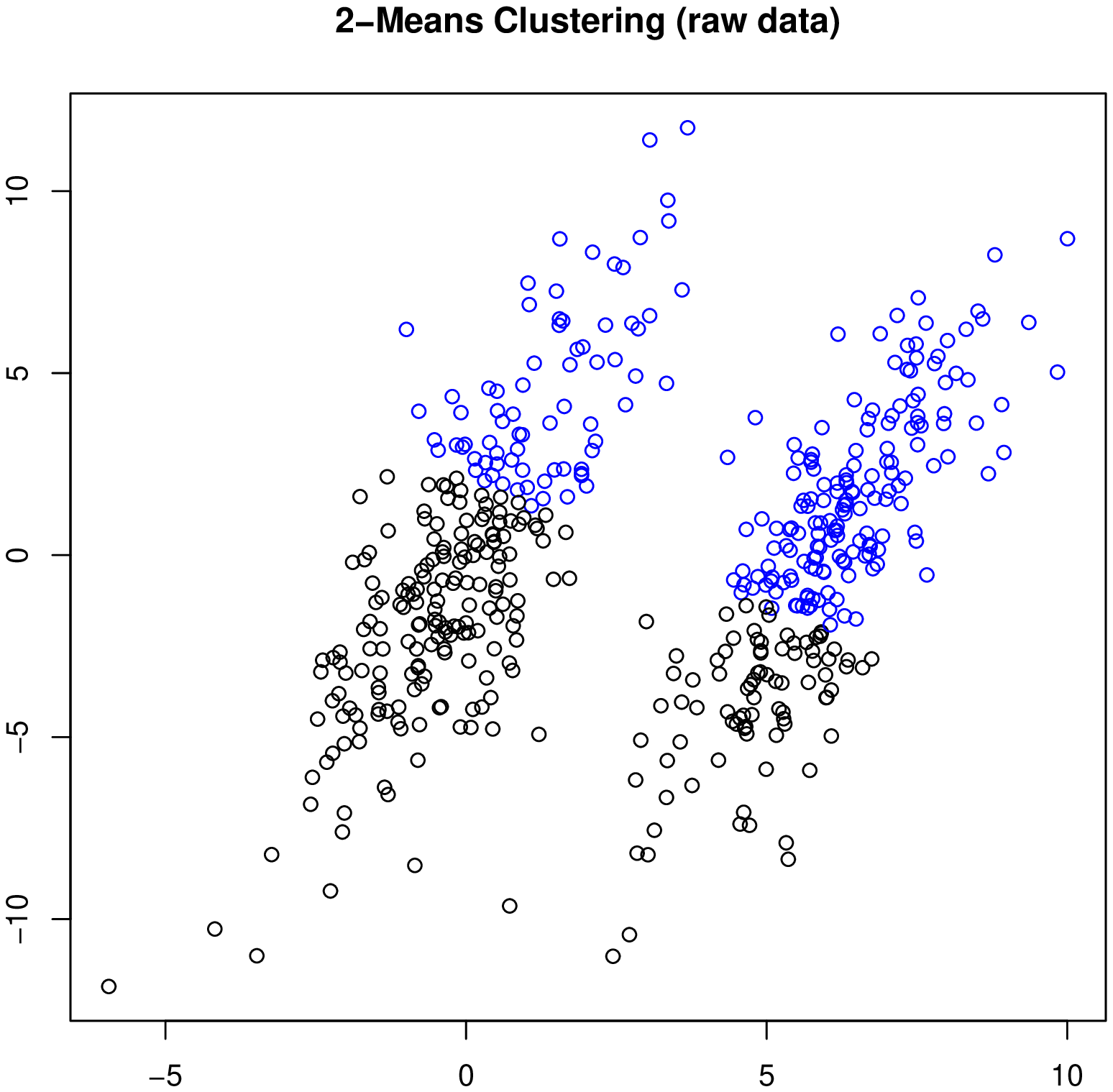}&
      \includegraphics[width=.5\textwidth]{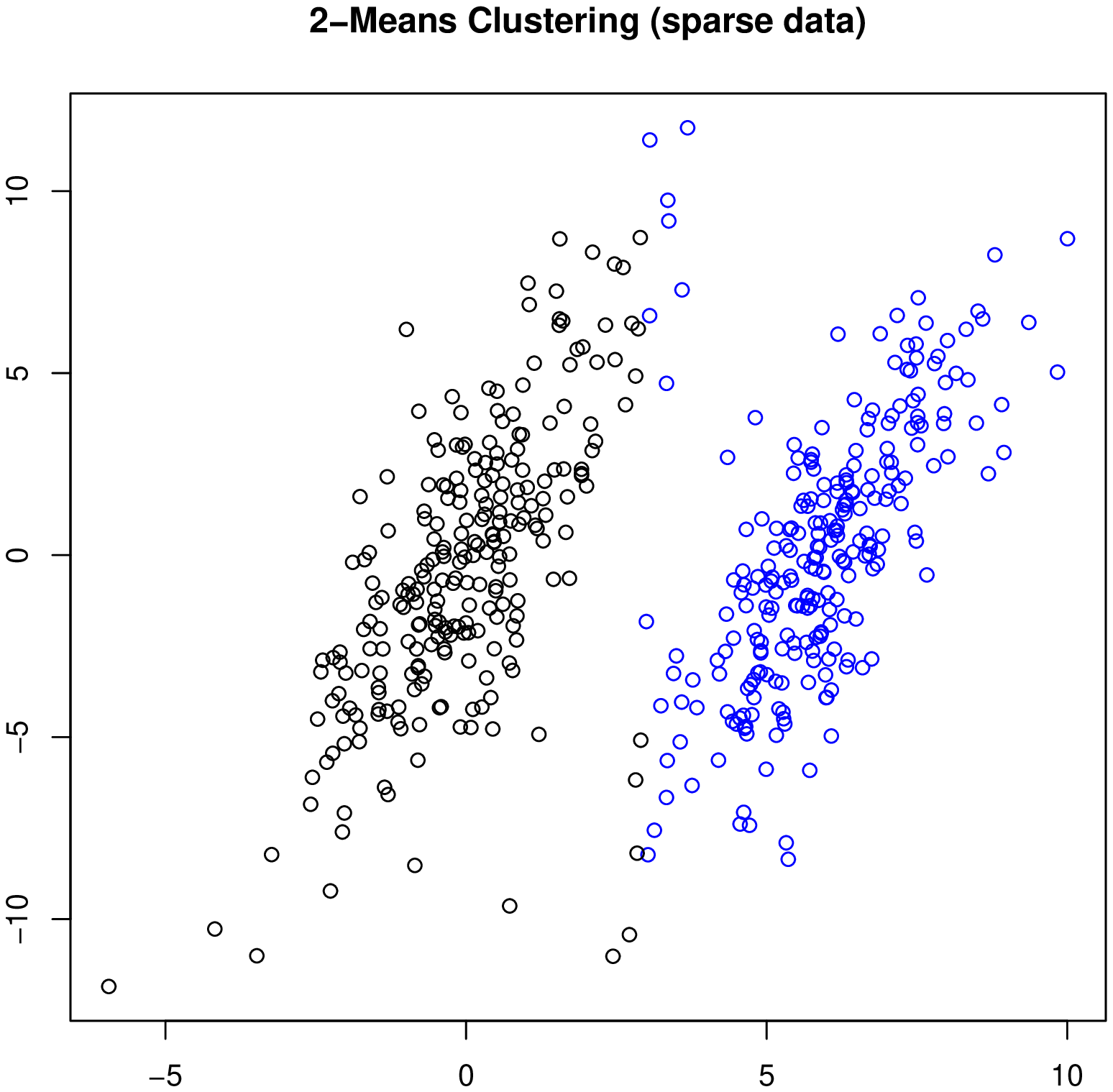}
      \end{array}$}
  \caption[Sparse clustering example]{In the above figure, there are
    two clusters such that the cluster means differ with respect to
    $x$ but not with respect to $y$. If 2-means clustering is applied
    to both $x$ and $y$, then it fails to identify the correct
    clusters, but 2-means clustering produces satisfactory results
    when applied only to $x$.}
  \label{F:sparse_clust}
\end{figure}

To overcome this difficulty, \citet{GB13} propose a method called
``supervised sparse clustering,'' which is a modification of the
``sparse clustering'' method of \citet{WT10}. Sparse clustering is an
unsupervised clustering method that is useful when the clusters differ
with respect to only a subset of the features. See Figure
\ref{F:sparse_clust} for an example of a data set where sparse
clustering produces better results than traditional k-means
clustering. In this (two-dimensional) example, the clusters differ
with respect to $x$ but not with respect to $y$. Applying 2-means
clustering to both $x$ and $y$ results produces inaccurate results,
but applying 2-means clustering only to $x$ identifies the correct
clusters.

The following is a brief description of the sparse clustering
algorithm of \citet{WT10}: First, note that minimizing the k-means
objective function (\ref{E:km_obj_func}) is equivalent to maximizing
\[
\sum_{j=1}^p \left( \frac{1}{n} \sum_{i=1}^n \sum_{i'=1}^n
  (x_{ij}-x_{i'j})^2 - \sum_{k=1}^K \frac{1}{n_k} \sum_{C_i=k}
  \sum_{C_{i'}=k} (x_{ij}-x_{i'j})^2 \right)
\]
Here each $x_{ij}$ is an observation from a data set with $n$
observations and $p$ features that is partitioned into $K$ clusters,
where $C_i=k$ if and only if observation $i$ belongs to cluster $k$
and $n_k$ is the number of observations in cluster $k$. Then the
sparse clustering algorithm seeks to identify weights $w_1, w_2,
\ldots, w_p$ for each feature to maximize
\begin{equation} \label{E:skm_obj_func}
  \sum_{j=1}^p \left[ w_j \left( \frac{1}{n} \sum_{i=1}^n
      \sum_{i'=1}^n (x_{ij}-x_{i'j})^2 - \sum_{k=1}^K \frac{1}{n_k}
      \sum_{C_i=k} \sum_{C_{i'}=k} (x_{ij}-x_{i'j})^2 \right) \right]
\end{equation}
subject to the constraints that $\sum_{j=1}^p w_j^2=1$, $\sum_{j=1}^p
|w_j| \leq s$, and $w_j \geq 0$ for all $j$. The variable $s$ is a
tuning parameter. As $s$ increases, the number of nonzero $w_j$'s will
increase. Thus, by choosing an appropriate value of $s$, the
clustering will be performed using only a subset of the features (the
features for which $w_j>0$). Note that sparse clustering imposes an
$L_1$ penalty on the feature weights, which is similar to the
$L_1$ penalty imposed on the regression coefficients in lasso
regression \citep{rT96} (which also causes an increasing number of
coefficients to be equal to 0 as the value of the tuning parameter
changes). See \citet{WT10} for a more detailed description of the
sparse clustering algorithm, including a method for choosing the
tuning parameter $s$. In particular, \citet{WT10} show that this
sparse clustering method tends to produce better results than several
previously published methods for reducing the dimension of a data set
prior to clustering, such as clustering on PCA scores \citep{GC02,
jL03}.

\citet{WT10} maximize (\ref{E:skm_obj_func}) by using an algorithm
that sets $w_j=1/\sqrt{n}$ at the beginning of the procedure and then
updates the $w_j$'s iteratively. The supervised sparse clustering
method of \citet{GB13} is similar to sparse clustering but chooses the
initial feature weights as follows:
\begin{enumerate}
\item For each feature in the data set, calculate a test statistic
  $T_j$ for testing the null hypothesis of no association between the
  $j$th feature and the outcome variable.
\item Choose a threshold $M$, and define the initial weights $w_1,
  w_2, \ldots, w_p$ as follows:
  \[
  w_j = \begin{cases}
    1/\sqrt{m} &\text{if $|T_j|>M$} \\
    0 &\text{if $|T_j| \leq M$}
  \end{cases}
  \]
  where $m$ is the number of features such that $|T_j|>M$.
\end{enumerate}
In other words, rather than giving equal initial weights to all the
features in the data set, supervised sparse clustering gives equal
initial weights to the features most strongly associated with the
outcome variable and an initial weight of 0 to all other
features. \citet{GB13} show that this modification of sparse
clustering is more likely to identify clusters that are associated
with an outcome variable.

Note that supervised sparse clustering is similar to several other
semi-supervised clustering methods. The method for choosing the
initial cluster weights is analogous to the method for choosing the
features in the supervised clustering algorithm of \citet{BT04}.
Indeed, the first step of the supervised sparse clustering algorithm
applies k-means clustering to the features most strongly associated
with the outcome variable, which is identical to the supervised
clustering method. The difference is that supervised sparse clustering
updates the feature weights after identifying the initial set of
clusters and iterates the procedure until convergence. \citet{GB13}
show that this procedure can produce better results than supervised
clustering in some situations, particularly when the outcome variable
is only weakly associated with the clusters. The supervised sparse
clustering procedure is also similar to the seeded k-means clustering
algorithm of \citet{sB02} since it uses the known outcome data to
``seed'' the initial step of the sparse clustering method and then
iterates the remainder of the sparse clustering algorithm without
further consideration of the outcome variable.

\begin{conclusion}
There has been considerable methodological research activity in the
area of semi-supervised clustering (particularly constrained
clustering) in the past decade. There now exists numerous methods for
performing constrained clustering (including the special case of
partially labeled data) that can be applied to a wide variety of
different data sets. In particular, several methods have been
developed for the special case of clustering genes in DNA microarray
data, where biological information often exists about the
relationships between some subset of the genes.

Nevertheless, there are several important unanswered questions in the
area of semi-supervised clustering. Although many algorithms exist for
performing constrained clustering, there does not appear to be
extensive research comparing the performance of the various
algorithms (either in terms of running time or in terms of their
ability to identify clusters correctly). Thus, users of these methods
may be uncertain about which method should be applied to a given data
set given the large number of options. Also, in the important special
case of genetic data, most existing research has focused on clustering
data from DNA microarrays. One might also wish to identify gene
clusters based on other types of modern high-throughput genetic data,
including data from genome-wide association studies, RNA-Seq, or
next-generation DNA sequencing. There is a need for semi-supervised
clustering methods that can be applied to these other types of genetic
data sets. Finally, as noted earlier, the problem of identifying
clusters associated with an outcome variable has not been studied
extensively in the literature. Only a handful of methods currently
exist. Development of new methods for this problem is another
potential area for future research.
\end{conclusion}

\section{Acknowledgments}
This work was partially supported by NIEHS grant P30ES010126 and NCATS
grant UL1RR025747. We thank the two anonymous reviewers for their
helpful suggestions.

\bibliographystyle{wires}
\bibliography{bibliography}

\begin{thebibliography}{70}
\providecommand{\natexlab}[1]{#1}
\expandafter\ifx\csname urlstyle\endcsname\relax
  \providecommand{\doi}[1]{doi:\discretionary{}{}{}#1}\else
  \providecommand{\doi}{doi:\discretionary{}{}{}\begingroup
  \urlstyle{rm}\Url}\fi

\bibitem[{Hastie et~al.(2009)Hastie, Tibshirani, and Friedman}]{HTF09}
Hastie T, Tibshirani R, Friedman JH.
\newblock The elements of statistical learning: data mining, inference, and
  prediction.
\newblock Springer Series in Statistics. Springer, New York, NY, 2009, 2nd
  edition.

\bibitem[{Gordon(1999)}]{aG99}
Gordon AD.
\newblock Classification.
\newblock Monographs on Statistics and Applied Probability. Chapman \& Hall,
  1999, 2nd edition.

\bibitem[{Blum and Mitchell(1998)}]{BM98}
Blum A, Mitchell T.
\newblock Combining labeled and unlabeled data with co-training.
\newblock In Proceedings of the 11th Annual Conference on Computational
  Learning Theory. 1998 92--100.

\bibitem[{Joachims(1999)}]{tJ99}
Joachims T.
\newblock Transductive inference for text classification using support vector
  machines.
\newblock In Proceedings of the 16th International Conference on Machine
  Learning (ICML-1999). 1999 200--209.

\bibitem[{Nigam et~al.(2000)Nigam, Mccallum, Thrun, and Mitchell}]{kN00}
Nigam K, Mccallum AK, Thrun S, Mitchell T.
\newblock Text classification from labeled and unlabeled documents using
  {E}{M}.
\newblock Machine Learning 2000.
\newblock 39:103--134.

\bibitem[{Basu et~al.(2004{\natexlab{a}})Basu, Bilenko, and Mooney}]{sB04b}
Basu S, Bilenko M, Mooney R.
\newblock A probabilistic framework for semi-supervised clustering.
\newblock In Proceedings of the 10th ACM SIGKDD International Conference on
  Knowledge Discovery and Data Mining. ACM, 2004{\natexlab{a}} 59--68.

\bibitem[{Forgy(1965)}]{eF65}
Forgy EW.
\newblock Cluster analysis of multivariate data: efficiency versus
  interpretability of classifications.
\newblock Biometrics 1965.
\newblock 21:768--769.

\bibitem[{MacQueen(1967)}]{jM67}
MacQueen J.
\newblock Some methods for classification and analysis of multivariate
  observations.
\newblock In Proceedings of the 5th Berkeley Symposium on Mathematical
  Statistics and Probability, edited by Le~Cam LM, Neyman J. University of
  California Press, Berkeley, CA, 1967, volume~1 281--297.

\bibitem[{Hartigan and Wong(1979)}]{HW79}
Hartigan JA, Wong MA.
\newblock Algorithm {A}{S} 136: A k-means clustering algorithm.
\newblock Journal of the Royal Statistical Society. Series C (Applied
  Statistics) 1979.
\newblock 28(1):100--108.

\bibitem[{Lloyd(1982)}]{sL82}
Lloyd S.
\newblock Least squares quantization in {P}{C}{M}.
\newblock IEEE Transactions on Information Theory 1982.
\newblock 28(2):129--137.

\bibitem[{Tibshirani et~al.(2001)Tibshirani, Walther, and Hastie}]{TWH01}
Tibshirani R, Walther G, Hastie T.
\newblock Estimating the number of clusters in a data set via the gap
  statistic.
\newblock Journal of the Royal Statistical Society: Series B (Statistical
  Methodology) 2001.
\newblock 63(2):411--423.
\newblock \doi{10.1111/1467-9868.00293}.

\bibitem[{Milligan and Cooper(1985)}]{MC85}
Milligan G, Cooper M.
\newblock An examination of procedures for determining the number of clusters
  in a data set.
\newblock Psychometrika 1985.
\newblock 50(2):159--179.
\newblock \doi{10.1007/BF02294245}.

\bibitem[{Sugar and James(2003)}]{SJ03}
Sugar CA, James GM.
\newblock Finding the number of clusters in a dataset.
\newblock Journal of the American Statistical Association 2003.
\newblock 98(463):750--763.
\newblock \doi{10.1198/016214503000000666}.

\bibitem[{Tibshirani and Walther(2005)}]{TW05}
Tibshirani R, Walther G.
\newblock Cluster validation by prediction strength.
\newblock Journal of Computational and Graphical Statistics 2005.
\newblock 14(3):511--528.
\newblock \doi{10.1198/106186005X59243}.

\bibitem[{Eisen et~al.(1998)Eisen, Spellman, Brown, and Botstein}]{mE98}
Eisen MB, Spellman PT, Brown PO, Botstein D.
\newblock Cluster analysis and display of genome-wide expression patterns.
\newblock Proceedings of the National Academy of Sciences 1998.
\newblock 95(25):14863--14868.

\bibitem[{Basu et~al.(2002)Basu, Banerjee, and Mooney}]{sB02}
Basu S, Banerjee A, Mooney R.
\newblock Semi-supervised clustering by seeding.
\newblock In Proceedings of the 19th International Conference on Machine
  Learning (ICML-2002). 2002 19--26.

\bibitem[{{Gaynor} and {Bair}(2013)}]{GB13}
{Gaynor} S, {Bair} E.
\newblock {Identification of biologically relevant subtypes via preweighted
  sparse clustering}.
\newblock ArXiv e-prints 2013.
\newblock arXiv:1304.3760.
\newblock {h}ttp://arxiv.org/abs/1304.3760.

\bibitem[{Brown et~al.(2000)Brown, Grundy, Lin, Cristianini, Sugnet, Furey,
  Ares, and Haussler}]{mB00}
Brown MPS, Grundy WN, Lin D, Cristianini N, Sugnet CW, Furey TS, Ares M,
  Haussler D.
\newblock Knowledge-based analysis of microarray gene expression data by using
  support vector machines.
\newblock Proceedings of the National Academy of Sciences 2000.
\newblock 97(1):262--267.
\newblock \doi{10.1073/pnas.97.1.262}.

\bibitem[{Mateos et~al.(2002)Mateos, Dopazo, Jansen, Tu, Gerstein, and
  Stolovitzky}]{aM02}
Mateos A, Dopazo J, Jansen R, Tu Y, Gerstein M, Stolovitzky G.
\newblock Systematic learning of gene functional classes from dna array
  expression data by using multilayer perceptrons.
\newblock Genome Research 2002.
\newblock 12(11):1703--1715.
\newblock \doi{10.1101/gr.192502}.

\bibitem[{Cheng et~al.(2004)Cheng, Cline, Martin, Finkelstein, Awad, Kulp, and
  Siani-Rose}]{jC04}
Cheng J, Cline M, Martin J, Finkelstein D, Awad T, Kulp D, Siani-Rose MA.
\newblock A knowledge-based clustering algorithm driven by gene ontology.
\newblock Journal of Biopharmaceutical Statistics 2004.
\newblock 14(3):687--700.
\newblock \doi{10.1081/BIP-200025659}.

\bibitem[{Qu and Xu(2004)}]{QX04}
Qu Y, Xu S.
\newblock Supervised cluster analysis for microarray data based on multivariate
  gaussian mixture.
\newblock Bioinformatics 2004.
\newblock 20(12):1905--1913.
\newblock \doi{10.1093/bioinformatics/bth177}.

\bibitem[{Fang et~al.(2006)Fang, Yang, Li, Luo, Liu et~al.}]{zF06}
Fang Z, Yang J, Li Y, Luo Q, Liu L, et~al.
\newblock Knowledge guided analysis of microarray data.
\newblock Journal of Biomedical Informatics 2006.
\newblock 39(4):401--411.

\bibitem[{Huang and Pan(2006)}]{HP06}
Huang D, Pan W.
\newblock Incorporating biological knowledge into distance-based clustering
  analysis of microarray gene expression data.
\newblock Bioinformatics 2006.
\newblock 22(10):1259--1268.
\newblock \doi{10.1093/bioinformatics/btl065}.

\bibitem[{Brameier and Wiuf(2007)}]{BW07}
Brameier M, Wiuf C.
\newblock Co-clustering and visualization of gene expression data and gene
  ontology terms for saccharomyces cerevisiae using self-organizing maps.
\newblock Journal of Biomedical Informatics 2007.
\newblock 40(2):160--173.
\newblock \doi{10.1016/j.jbi.2006.05.001}.

\bibitem[{Chopra et~al.(2008)Chopra, Kang, Yang, Cho, Kim, and Lee}]{pC08}
Chopra P, Kang J, Yang J, Cho H, Kim H, Lee MG.
\newblock Microarray data mining using landmark gene-guided clustering.
\newblock BMC Bioinformatics 2008.
\newblock 9(1):92.
\newblock \doi{10.1186/1471-2105-9-92}.

\bibitem[{Tari et~al.(2009)Tari, Baral, and Kim}]{TBK09}
Tari L, Baral C, Kim S.
\newblock Fuzzy c-means clustering with prior biological knowledge.
\newblock Journal of Biomedical Informatics 2009.
\newblock 42(1):74--81.
\newblock \doi{10.1016/j.jbi.2008.05.009}.

\bibitem[{Basu et~al.(2009)Basu, Davidson, and Wagstaff}]{BDW09}
Basu S, Davidson I, Wagstaff K.
\newblock Constrained Clustering: Advances in Algorithms, Theory, and
  Applications.
\newblock Chapman \& Hall/CRC Data Mining and Knowledge Discovery Series. CRC
  Press, Boca Raton, FL, 2009.

\bibitem[{Wagstaff et~al.(2001)Wagstaff, Cardie, Rogers, and
  Schr{\"o}dl}]{kW01}
Wagstaff K, Cardie C, Rogers S, Schr{\"o}dl S.
\newblock Constrained k-means clustering with background knowledge.
\newblock In Proceedings of the 18th International Conference on Machine
  Learning (ICML-2001). 2001 577--584.

\bibitem[{Basu et~al.(2004{\natexlab{b}})Basu, Banerjee, and Mooney}]{sB04}
Basu S, Banerjee A, Mooney R.
\newblock Active semi-supervision for pairwise constrained clustering.
\newblock In Proceedings of the 4th SIAM International Conference on Data
  Mining (SDM-2004). 2004{\natexlab{b}} 333--344.

\bibitem[{Bilenko et~al.(2004)Bilenko, Basu, and Mooney}]{BBM04}
Bilenko M, Basu S, Mooney R.
\newblock Integrating constraints and metric learning in semi-supervised
  clustering.
\newblock In Proceedings of the 21st International Conference on Machine
  learning (ICML-2004). 2004 81--88.

\bibitem[{Klein et~al.(2002)Klein, Kamvar, and Manning}]{KKM02}
Klein D, Kamvar S, Manning C.
\newblock From instance-level constraints to space-level constraints: Making
  the most of prior knowledge in data clustering.
\newblock In Proceedings of the 19th International Conference on Machine
  Learning (ICML-2002). 2002 307--314.

\bibitem[{Bilenko and Mooney(2003)}]{BM03}
Bilenko M, Mooney R.
\newblock Adaptive duplicate detection using learnable string similarity
  measures.
\newblock In Proceedings of the 9th ACM SIGKDD International Conference on
  Knowledge Discovery and Data Mining. 2003 39--48.

\bibitem[{Xing et~al.(2003)Xing, Ng, Jordan, and Russell}]{eX03}
Xing E, Ng A, Jordan M, Russell S.
\newblock Distance metric learning, with application to clustering with
  side-information.
\newblock In Advances in Neural Information Processing Systems 15. 2003
  505--512.

\bibitem[{Bar-Hillel et~al.(2003)Bar-Hillel, Hertz, Shental, and
  Weinshall}]{aB03}
Bar-Hillel A, Hertz T, Shental N, Weinshall D.
\newblock Learning distance functions using equivalence relations.
\newblock In Proceedings of the 20th International Conference on Machine
  learning (ICML-2003). 2003 11--18.

\bibitem[{Kamvar et~al.(2003)Kamvar, Klein, and Manning}]{KKM03}
Kamvar S, Klein D, Manning C.
\newblock Spectral learning.
\newblock In Proceedings of the 17th International Joint Conference of
  Artificial Intelligence. 2003 561--566.

\bibitem[{Chang and Yeung(2004)}]{CY04}
Chang H, Yeung DY.
\newblock Locally linear metric adaptation for semi-supervised clustering.
\newblock In Proceedings of the 21st International Conference on Machine
  learning (ICML-2004). 2004 153--160.

\bibitem[{Lange et~al.(2005)Lange, Law, Jain, and Buhmann}]{tL05}
Lange T, Law M, Jain A, Buhmann J.
\newblock Learning with constrained and unlabelled data.
\newblock In Proceedings of the IEEE Conference on Computer Vision and Pattern
  Recognition (CVPR 2005). 2005 731--738.

\bibitem[{Handl and Knowles(2006)}]{HK06}
Handl J, Knowles J.
\newblock On semi-supervised clustering via multiobjective optimization.
\newblock In Proceedings of the 8th Annual Conference on Genetic and
  Evolutionary Computation (GECCO 2006). 2006 1465--1472.

\bibitem[{Li et~al.(2007)Li, Ding, and Jordan}]{LDJ07}
Li T, Ding C, Jordan M.
\newblock Solving consensus and semi-supervised clustering problems using
  nonnegative matrix factorization.
\newblock In Proceedings of the 7th IEEE International Conference on Data
  Mining (ICDM 2007). 2007 577--582.

\bibitem[{Xiang et~al.(2008)Xiang, Nie, and Zhang}]{XNZ08}
Xiang S, Nie F, Zhang C.
\newblock Learning a {M}ahalanobis distance metric for data clustering and
  classification.
\newblock Pattern Recognition 2008.
\newblock 41(12):3600--3612.
\newblock \doi{10.1016/j.patcog.2008.05.018}.

\bibitem[{Wang et~al.(2008)Wang, Li, and Zhang}]{WLZ08}
Wang F, Li T, Zhang C.
\newblock Semi-supervised clustering via matrix factorization.
\newblock In Proceedings of the 8th SIAM International Conference on Data
  Mining (SDM-2008). 2008 1--12.

\bibitem[{Cohn et~al.(2009)Cohn, Caruana, and McCallum}]{CCM09}
Cohn D, Caruana R, McCallum A.
\newblock Semi-supervised clustering with user feedback.
\newblock In Constrained Clustering: Advances in Algorithms, Theory, and
  Applications, edited by Basu S, Davidson I, Wagstaff K, CRC Press, Boca
  Raton, FL, 2009, Chapman \& Hall/CRC Data Mining and Knowledge Discovery
  Series, chapter~2, 17--31.

\bibitem[{Yin et~al.(2010)Yin, Chen, Hu, and Zhang}]{xY10}
Yin X, Chen S, Hu E, Zhang D.
\newblock Semi-supervised clustering with metric learning: An adaptive kernel
  method.
\newblock Pattern Recognition 2010.
\newblock 43(4):1320--1333.
\newblock \doi{10.1016/j.patcog.2009.11.005}.

\bibitem[{Davidson and Ravi(2005{\natexlab{a}})}]{DR05}
Davidson I, Ravi S.
\newblock Clustering with constraints: Feasibility issues and the k-means
  algorithm.
\newblock In Proceedings of the 5th SIAM International Conference on Data
  Mining (SDM-2005). 2005{\natexlab{a}} 138--149.

\bibitem[{Davidson and Ravi(2005{\natexlab{b}})}]{DR05b}
Davidson I, Ravi S.
\newblock Agglomerative hierarchical clustering with constraints: Theoretical
  and empirical results.
\newblock In Knowledge Discovery in Databases (KDD 2005). 2005{\natexlab{b}}
  59--70.

\bibitem[{Law et~al.(2005)Law, Topchy, and Jain}]{LTJ05}
Law M, Topchy A, Jain A.
\newblock Model-based clustering with probabilistic constraints.
\newblock In Proceedings of the 5th SIAM International Conference on Data
  Mining (SDM-2005). 2005 641--645.

\bibitem[{Lu and Leen(2005)}]{LL05}
Lu Z, Leen T.
\newblock Semi-supervised learning with penalized probabilistic clustering.
\newblock In Advances in Neural Information Processing Systems 17. 2005
  849--856.

\bibitem[{Tang et~al.(2007)Tang, Xiong, Zhong, and Wu}]{wT07}
Tang W, Xiong H, Zhong S, Wu J.
\newblock Enhancing semi-supervised clustering: a feature projection
  perspective.
\newblock In Proceedings of the 13th ACM SIGKDD International Conference on
  Knowledge Discovery and Data Mining. 2007 707--716.

\bibitem[{Kulis et~al.(2009)Kulis, Basu, Dhillon, and Mooney}]{bK09}
Kulis B, Basu S, Dhillon I, Mooney R.
\newblock Semi-supervised graph clustering: a kernel approach.
\newblock Machine Learning 2009.
\newblock 74:1--22.
\newblock \doi{10.1007/s10994-008-5084-4}.

\bibitem[{Yoshida and Okatani(2010)}]{YO10}
Yoshida T, Okatani K.
\newblock A graph-based projection approach for semi-supervised clustering.
\newblock In Knowledge Management and Acquisition for Smart Systems and
  Services, edited by Kang BH, Richards D, Springer-Verlag, Berlin, Germany,
  2010, Lecture Notes in Computer Science, 1--13.

\bibitem[{Greene and Cunningham(2007)}]{GC07}
Greene D, Cunningham P.
\newblock Constraint selection by committee: an ensemble approach to
  identifying informative constraints for semi-supervised clustering.
\newblock In Proceedings of the 18th European Conf. on Machine Learning (ECML
  2007). 2007 140--151.

\bibitem[{Mallapragada et~al.(2008)Mallapragada, Jin, and Jain}]{pM08}
Mallapragada P, Jin R, Jain A.
\newblock Active query selection for semi-supervised clustering.
\newblock In 19th International Conference on Pattern Recognition (ICPR 2008).
  IEEE, 2008 1--4.

\bibitem[{Zheng and Li(2011)}]{ZL11}
Zheng L, Li T.
\newblock Semi-supervised hierarchical clustering.
\newblock In Proceedings of the 11th IEEE International Conference on Data
  Mining (ICDM 2011). 2011 982--991.
\newblock \doi{10.1109/ICDM.2011.130}.

\bibitem[{Miyamoto and Terami(2010)}]{MT10}
Miyamoto S, Terami A.
\newblock Semi-supervised agglomerative hierarchical clustering algorithms with
  pairwise constraints.
\newblock In Proceedings of the 2010 IEEE International Conference on Fuzzy
  Systems (FUZZ 2010). 2010 1--6.
\newblock \doi{10.1109/FUZZY.2010.5584625}.

\bibitem[{Davidson and Ravi(2009)}]{DR09}
Davidson I, Ravi S.
\newblock Using instance-level constraints in agglomerative hierarchical
  clustering: theoretical and empirical results.
\newblock Data Mining and Knowledge Discovery 2009.
\newblock 18(2):257--282.
\newblock \doi{10.1007/s10618-008-0103-4}.

\bibitem[{Miyamoto and Terami(2011)}]{MT11}
Miyamoto S, Terami A.
\newblock Constrained agglomerative hierarchical clustering algorithms with
  penalties.
\newblock In Proceedings of the 2011 IEEE International Conference on Fuzzy
  Systems (FUZZ 2011). 2011 422--427.
\newblock \doi{10.1109/FUZZY.2011.6007351}.

\bibitem[{Bade and Nurnberger(2006)}]{BN06}
Bade K, Nurnberger A.
\newblock Personalized hierarchical clustering.
\newblock In Proceedings of the 2006 IEEE/WIC/ACM International Conference on
  Web Intelligence (WI 2006). 2006 181--187.
\newblock \doi{10.1109/WI.2006.131}.

\bibitem[{Zhao and Qi(2010)}]{ZQ10}
Zhao H, Qi Z.
\newblock Hierarchical agglomerative clustering with ordering constraints.
\newblock In Proceedings of the 3rd International Conference on Knowledge
  Discovery and Data Mining (WKDD 2010). 2010 195--199.
\newblock \doi{10.1109/WKDD.2010.123}.

\bibitem[{Hamasuna et~al.(2010)Hamasuna, Endo, and Miyamoto}]{YYS10}
Hamasuna Y, Endo Y, Miyamoto S.
\newblock Semi-supervised agglomerative hierarchical clustering using
  clusterwise tolerance based pairwise constraints.
\newblock In Proceedings of the 7th International Conference on Modeling
  Decision for Artificial Intelligence (MDAI 2010). 2010 152--162.

\bibitem[{Hamasuna et~al.(2011)Hamasuna, Endo, and Miyamoto}]{YYS11}
Hamasuna Y, Endo Y, Miyamoto S.
\newblock Semi-supervised agglomerative hierarchical clustering with ward
  method using clusterwise tolerance.
\newblock In Proceedings of the 8th International Conference on Modeling
  Decision for Artificial Intelligence (MDAI 2011). 2011 103--113.

\bibitem[{Hamasuna et~al.(2012)Hamasuna, Endo, and Miyamoto}]{YYS12}
Hamasuna Y, Endo Y, Miyamoto S.
\newblock On agglomerative hierarchical clustering using clusterwise tolerance
  based pairwise constraints.
\newblock Journal of Advanced Computational Intelligence and Intelligent
  Informatics 2012.
\newblock 16(1):174--179.

\bibitem[{Bair and Tibshirani(2004)}]{BT04}
Bair E, Tibshirani R.
\newblock Semi-supervised methods to predict patient survival from gene
  expression data.
\newblock PLoS Biol 2004.
\newblock 2(4):e108.
\newblock \doi{10.1371/journal.pbio.0020108}.

\bibitem[{Nowak and Tibshirani(2008)}]{NT08}
Nowak G, Tibshirani R.
\newblock Complementary hierarchical clustering.
\newblock Biostatistics 2008.
\newblock 9(3):467--483.
\newblock \doi{10.1093/biostatistics/kxm046}.

\bibitem[{Bullinger et~al.(2004)Bullinger, D\"ohner, Bair, Fr\"ohling, Schlenk,
  Tibshirani, D\"ohner, and Pollack}]{lB04}
Bullinger L, D\"ohner K, Bair E, Fr\"ohling S, Schlenk R, Tibshirani R,
  D\"ohner H, Pollack JR.
\newblock Gene expression profiling identifies new subclasses and improves
  outcome prediction in adult myeloid leukemia.
\newblock The New England Journal of Medicine 2004.
\newblock 350:1605--1616.

\bibitem[{Koestler et~al.(2010)Koestler, Marsit, Christensen, Karagas, Bueno,
  Sugarbaker, Kelsey, and Houseman}]{dK10}
Koestler DC, Marsit CJ, Christensen BC, Karagas MR, Bueno R, Sugarbaker DJ,
  Kelsey KT, Houseman EA.
\newblock Semi-supervised recursively partitioned mixture models for
  identifying cancer subtypes.
\newblock Bioinformatics 2010.
\newblock 26(20):2578--2585.
\newblock \doi{10.1093/bioinformatics/btq470}.

\bibitem[{Houseman et~al.(2008)Houseman, Christensen, Yeh, Marsit, Karagas,
  Wrensch, Nelson, Wiemels, Zheng, Wiencke et~al.}]{eH08}
Houseman EA, Christensen B, Yeh RF, Marsit C, Karagas M, Wrensch M, Nelson H,
  Wiemels J, Zheng S, Wiencke J, et~al.
\newblock Model-based clustering of dna methylation array data: a
  recursive-partitioning algorithm for high-dimensional data arising as a
  mixture of beta distributions.
\newblock BMC Bioinformatics 2008.
\newblock 9(1):365.
\newblock \doi{10.1186/1471-2105-9-365}.

\bibitem[{Witten and Tibshirani(2010)}]{WT10}
Witten DM, Tibshirani R.
\newblock A framework for feature selection in clustering.
\newblock Journal of the American Statistical Association 2010.
\newblock 105(490):713--726.
\newblock \doi{10.1198/jasa.2010.tm09415}.

\bibitem[{Tibshirani(1996)}]{rT96}
Tibshirani R.
\newblock Regression shrinkage and selection via the lasso.
\newblock Journal of the Royal Statistical Society. Series B (Methodological)
  1996.
\newblock 58(1):267--288.

\bibitem[{Ghosh and Chinnaiyan(2002)}]{GC02}
Ghosh D, Chinnaiyan AM.
\newblock Mixture modelling of gene expression data from microarray
  experiments.
\newblock Bioinformatics 2002.
\newblock 18(2):275--286.
\newblock \doi{10.1093/bioinformatics/18.2.275}.

\bibitem[{Liu et~al.(2003)Liu, Zhang, Palumbo, and Lawrence}]{jL03}
Liu JS, Zhang JL, Palumbo MJ, Lawrence CE.
\newblock Bayesian clustering with variable and transformation selections.
\newblock In Bayesian Statistics 7: Proceedings of the Seventh Valencia
  International Meeting. 2003 249--275.

\end{thebibliography}

\begin{crossref}
\cref{CSV0045}, \cref{CSA0081}, \cref{CSV0056}
\end{crossref}

\endAdvancedReview
\end{document}